\documentclass[10pt,twocolumn]{article}
\usepackage{graphicx}
\usepackage{geometry} 
\usepackage{color}
\begin{document}
\title{Social network markets: the influence of network structure when consumers face decisions over many similar choices }\author{Paul Ormerod\\Volterra Partners LLP, London \& University of Durham \and Bassel Tarbush\\Department of Economics, University of Oxford  \and R. Alexander Bentley\\Department of Archaeology and Anthropology, University of Bristol} \maketitle

\textbf{In social network markets, the act of consumer choice in these industries is governed not just by the set of incentives described by conventional consumer demand theory, but by the choices of others in which an individualÕs payoff is an explicit function of the actions of others. We observe two key empirical features of outcomes in social networked markets.  First, a highly right-skewed, non-Gaussian distribution of the number of times competing alternatives are selected at a point in time.  Second, there is turnover in the rankings of popularity over time. We show here that such outcomes can arise either when there is no alternative which exhibits inherent superiority in its attributes, or when agents find it very difficult to discern any differences in quality amongst the alternatives which are available so that it is as if no superiority exists.  These features appear to obtain, as a reasonable approximation, in many social network markets. We examine the impact of network structure on both the rank-size distribution of choices at a point in time, and on the life spans of the most popular choices.  We show that a key influence on outcomes is the extent to which the network follows a hierarchical structure.  It is the social network properties of the markets, the meso-level structure, which determine outcomes rather than the objective attributes of the products.}

\section{Introduction}
\label{intro}
Potts et al. (2008) argue that the analytical foundation for the cultural and creative industries is best provided by taking the perspective of an emergent market economy rather than an industrial one.  They describe these industries, and others which share their fundamental features, as `social network markets'.  

They suggest that ``the very act of consumer choice in these industries is governed not just by the set of incentives described by conventional consumer demand theory, but by the choices of others in which an individualÕs payoff is an explicit function of the actions of others.'' Examples are given by Arthur (1989), Brock and Durlauf (2001, 2010), De Vany and Walls (1996), Ormerod (1998, 2005, 2007), Kretschmer {\it et al.} (1999), Beck (2007) and Bentley and Ormerod (2009). These phenomena, which Schelling (1973) described generally as  ``binary decisions with externalities,''  clearly apply generally to the creative industries ({\it e.g.}, De Vany 2004, Potts 2006, Beck 2007).

A key question when considering how agents choose between alternatives in such networked markets is whether those alternatives which are superior to others in terms of their attributes come to the fore. In other words, are superior alternatives more likely to be chosen than ones which are in some sense inferior? This is obviously important in not just in modern cultural and creative markets, but in areas such as the diffusion of innovation and technology (for example, Antonelli, ed., 2011).  It also appears in such disparate contexts as foraging theory in early human societies (Winterhalder and Smith 2000), and in prehistoric choices of ceramic pottery (Neiman 1995).

Different approaches have been proposed in addressing this question. In the economics literature, the seminal paper of Akerlof (1970) showed that the inter-action between the heterogeneity of quality amongst the alternatives and asymmetric information amongst agents, so that some or even all agents lack complete information, can lead not only to superior alternatives being withdrawn from the market, but to the market ceasing to function at all. Simon (1955) argued that, in general, agents lack the processing capacity to select the optimal choice, even when, indeed especially when, complete information is available. Arthur (1989) considered agents who choose between two competing variants of a new technology; they lack the information which would enable them to distinguish between their qualities, and as a consequence the inferior technology can capture the entire market.

As Earls and Potts (2004) point out, the issue is not just that agents have incomplete or asymmetric information, but also that they have different and possibly better or more experienced choice rules. In other words, we might still expect to observe this behaviour even when there is not asymmetric or distributed information.  

Recent social network models from the network literature have focused on the issues of the dynamics of popularity and the process by which `superior' alternatives come to the fore.   A seminal contribution was made by Lieberman et al. (2005). They used an established evolutionary model to capture the process of selection and choice.  They showed that the probability that a new, better invention becomes adopted widely depends in a fundamental way on the social network structure. 

Hierarchical networks, for example, did a better job at ensuring that inherently superior options would be adopted across the population.  Diffused or random networks tended to minimise the advantage of any inherent superiority. In one extreme category of diffuse networks, the superior option had almost no advantage, whereas in the other limit of extremely hierarchical network structure, any superiority of a new invention was bound to be picked up and spread. 

In this paper, we examine the impact of network structure on the two key empirical features of social network markets, which we describe below.  We do so in contexts in which either there is no alternative which exhibits inherent superiority in its attributes, or in which agents find it very difficult to discern any differences in quality amongst the alternatives which are available so that it is as if no superiority exists.

In the marketing literature, a seminal analysis of such markets was carried out by Ehrenberg (1959), who considered consumer markets in which consumers effectively guessed from among the choices available.    He assumed that choice ``incidence tends to beÉ so irregular that it can be regarded as if random'', (Goodhart, Ehrenberg {\it et al.} 1984) and from this he predicted a negative binomial distribution of variant popularity.

Certainly, there is now a stupendous variety of choices available in many consumer markets. Consider, for example, what Beinhocker (2007) has said about the plethora of consumer choices for the average New Yorker: ``The Wal-Mart near JFK Airport has over 100,000 different items in stock, there are over200 television channels offered on cable TV, Barnes and Noble lists over 8 million titles, the local supermarket has 275 varieties of breakfast cereal, the typical department store offers 150 types of lipstick, and there are over 50,000 restaurants in New York City alone.''  At the stock keeping unit level (SKU), the level of product detail at which retailers specify their restocking orders, Beinhocker estimates that on a single day in New York, there are 10 billion (!) such choices available.   There may indeed be objective differences between the various offers, but in such numerous, minor and often incomprehensible ways that they exemplify what has come to be called `decision quicksand' (Sela and Berger 2011) or `decision fatigue' (Baumeister and Tierney 2011).

In such circumstances, the plethora of choice makes processing of the information in the classical rational economic sense effectively impossible.   Rosen (1981) analysed the economics of superstars. His model was generalised by Adler (1985, 2006) to examine the consumer side where having and sharing knowledge and information about artistic goods minimises searching costs (time), generating learning effects and positive externalities, even when the top few `superstars' are not necessarily objectively better than the rest of the alternatives.  

We therefore examine the impact of network structure on outcomes in social network markets in which we assume that none of the alternative offers which are available have any intrinsic superiority over any of the others.

We focus our attention on the impact of network structure on two key empirical outcomes of social network markets.  First, the popularity of different choices at a point in time follows a right-skewed, non-Gaussian distribution.  The music download experiment of Salganik et al. (2006), for example, shows how the impact of social networks dramatically increases the non-Gaussian features of the statistical distribution of the outcome of choice made about different alternatives.  A slightly irreverent but highly relevant example is given by Godart and Mears (2009).  They studied the Style.com show reports for Spring 2007 and found that designers used a total of 677 fashion models worldwide for their shows.   Over 75 per cent of the total who appeared, were in 5 shows or less.  Only 60 women in the entire modelling universe walked in more than 20 shows, with the market leader of the time, Coco Rocha, being featured in 55 shows.  Many aspirants, of course, featured in none.

Kahneman (2011, chapter 17) hypothesises that sales of competing alternatives tends to exhibit regression to the mean.  The heavily right skewed nature of outcomes in social network markets requires that this statement be qualified substantially.  In the long run, in social network markets, the process of evolution which drives them means that the sales of any product will fall to zero.  So the long-run mean is zero.  However, in the short-term -- and the illustrative example provided by Kahneman is one-step ahead projection -- the self-reinforcing nature of the process of choice means that products which are rising in popularity will in general exhibit the complete opposite of mean reversion.

The second key empirical feature is that there is turnover in the rankings of popularity over time.  Ormerod (2012) provides numerous examples of both these phenomena.

We examine the impact of network structure on these two features of social network markets.  We find that the network structure can have substantial effects on the extent of skewness in the distribution and on the speed of turnover in the rankings.

We consider four distinct large networks that match those studied by Lieberman et al. (2005) which, from the least hierarchical to the most hierarchical, were: (a) a square lattice, where agents copy other agents near them on a grid, (b) a fully connected network, where each agent can copy any other agent, (c) a ``meta-funnel'' network, where the structure funnels out from a central agent, and (d) a ``superstar'' network, where agents are grouped and a central agent is connected to all groups.

Section 2 describes the behavioural model of agent choice and the network structures examined.  Section 3 sets out the results, and section 4 offers a brief discussion and conclusion.

\section{Models and methods}
\label{sec:2}
Given that, by assumption, agents at any point in time make a choice amongst a large number of indistinguishable alternatives, the behavioural model of economics,  that of rational selection on the basis of objective information, is not relevant, even when it is modified to take into account imperfect and asymmetric information.

\subsection{Model}
\label{sec:2.1}
Given that, by assumption, agents at any point in time make a choice amongst a large number of indistinguishable alternatives, the behavioural model of economics, that of rational selection on the basis of objective information, faces challenges, even when it is modified to take into account imperfect and asymmetric information.  If rationality is defined as maximizing utility subject to constraints, but every possible good is effectively identical, then every good will be in the argmax of the utility, and therefore every good will be chosen with equal probability.  This is essentially the argument of Ehrenberg (1959).  Our evolutionary choice model, however, differs from this.  As a heuristic in such circumstances, agents essentially choose with a probability equal to the number of times any given alternative has been selected as a proportion of the total number of selections made across the agents to which the agent is connected.  They may, for example, regard other agents as having more information than they do, and hence copy their behaviour. 

In economics, Brock and Durlauf (1999, 2001, 2010), and Young (2009, 2011), for example, have explored the effects of social interactions and social diffusion in great depth.  Here we hope to contribute a complementary behavioural model of a different nature, developed from the cultural evolution literature, which has been shown to be capable of explaining empirical outcomes in a wide range of different contexts.   A majority of agents in a population copy one another in an effectively undirected manner, and the (small) minority innovate their own original behaviours.  Lieberman et al. (2005) used a similar copying model.   In terms of understanding the behaviour of this model, there are examples from biological sciences (Hahn 2008), marketing science (Goodhardt {\it et al.} 1984), and economics (Ijiri and Simon 1964).  Practical applications, with favourable comparison to real-world data, include baby names (Hahn and Bentley 2003), English words (Bentley 2008; Reali and Griffiths 2004), prehistoric pottery designs (Neiman 1995) and key features of linguistic evolution (Bentley {\it et al.} 2011a).  A generalisation of the model is given by Bentley {\it et al.} (2011b, 2011c).

This modelling approach differs from that based upon the concept of rational addiction with preferences which are learned and are inter--temporally dependent (for example, Becker and Murphy 1988, Britto and Barros 2005).  Agents are not required to learn preferences over time.  In the cultural evolution literature, the preferences of any given agent are not formed over time.  At any point in time, an agent makes a choice based simply on the choices made by others, with a small probability of random innovation in making their selection.  As noted above, this choice rule could be used because an agent might believe others to possess either superior information or superior decision rules.  Strong evidence for the effectiveness of copying as the basis for decisions is provide by the social learning tournament organised by Rendell {\it et al.} (2010). 
 There is also well documented psychological evidence as to why agents might make use of this decision rule, such as the desire to conform ({\it e.g.}, Asch (1955), Moscovici {\it et al.} 1969) or the influence of peer acceptance ({\it e.g.}, Christiakis and Fowler 2007; Ormerod and Wiltshire 2009; Rivera {\it et al.} 2010)

The model proceeds as follows: There are $N$ agents in a fixed network. There are initially $N$ possible choices, and we initialise the model with a unique and distinct choice being assigned to each agent. In each period, every agent can make a choice. With probability $1-\mu$ the agent selects a choice with a probability proportional to the number of agents that it is connected to that have made that choice in the previous period, and with probability $\mu$, the agent selects a new choice that is different from all previously available choices.  Considerable anthropological and socioeconomic evidence exists (e.g., Eerkens 2000; Diederen {\it et al.} 2003; Srinivasan and Mason 1986; Larsen 1961; Rogers 1962) on the plausible values for $\mu$ being no greater than 0.1.

\begin{figure}
\begin{center}
 \includegraphics[width=0.45\textwidth]{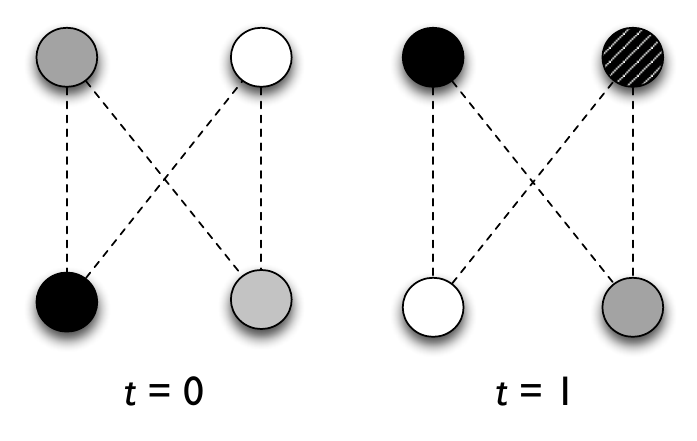}
\caption{A simple representation of the neutral model.  Shown are four individuals for two successive time steps.  At each time step, we refresh the population with new individuals, and each is given a new copy of a variant (represented by the shading within the circles).  Each variant is assigned a new value by either (a) copying a randomly-selected individual from the previous time step, with equal probability of choosing any individual, or (b) inventing a new variant (e.g. dotted shading) with probability $\mu$, which in this example is 0.25 (one our of the four individuals).}
\label{fig:1}       
\end{center}
\end{figure}

An intuitive understanding of the model is given in Fig.~\ref{fig:1}.

We initialise the model at $t = 0$ by letting each agent select its own choice, represented here by a colour.  So the agent in the top left selects `grey' for example.   When we move to $t =1$, agents have the choice of copying the choice of any agent to which it is connected or, with a small probability, making a completely new choice.  There are many possible configurations that can result from this at $t = 1$, and only one such configuration is represented in the figure. For example, the top-left agent switched from grey to black by copying the agent at the bottom right at time $t = 0$. This is possible since this agent is connected to a black node and a light grey one. On the other hand, it could not have switched to white. Similarly, the bottom-left agent switched from black to white, and the bottom-right agent switched from light grey to grey. Finally, the top-right agent switched from whited to striped.  In other words, this latter agent experimented,  by selecting a choice that had not previously been made.

We ran this model 500 independent times over each of four different large network structures (approximately 475 nodes), and for 2000 periods each time. The different network structures were chosen to resemble the network arrangements studied in Lieberman et al. (2005). We also varied the experimentation parameter $\mu$, allowing it to take the values 0, 0.25\%, 0.5\%, 0. 75\%,  1\% and 5\%.

For each run of the model, we recorded:
\begin{itemize}
\item The distribution of choices ranked by popularity
\item The life spans of the top 100 longest surviving choices
\item The life spans of the top 100 most popular choices (measured by the total number of times the choice was selected over the run of the model)
\item The number of choices that are counted among the top 100 longest survivors and the top 100 most popular
\end{itemize}

Since the model was run 500 independent times over each network, we obtained the averages over the 500 runs of the above statistics.

\subsection{Network structures}
\label{sec:2.2}
The four network structures are those examined in Lieberman {\it et al.}, to facilitate direct comparison between social network markets in which one of the alternatives is superior in its attributes to those of the others, and markets in which none of the alternatives is superior. 

The key difference between them which is relevant in this context is the extent to which they have a hierarchical structure.  In other words, the extent to which there is a small number of highly connected nodes which therefore have the potential to exercise a strong influence on the outcome.  Details of the graph theoretic properties of the networks are available from the authors, and here we set out a brief description.

\begin{figure*}
\begin{center}
\includegraphics[width=0.75\textwidth]{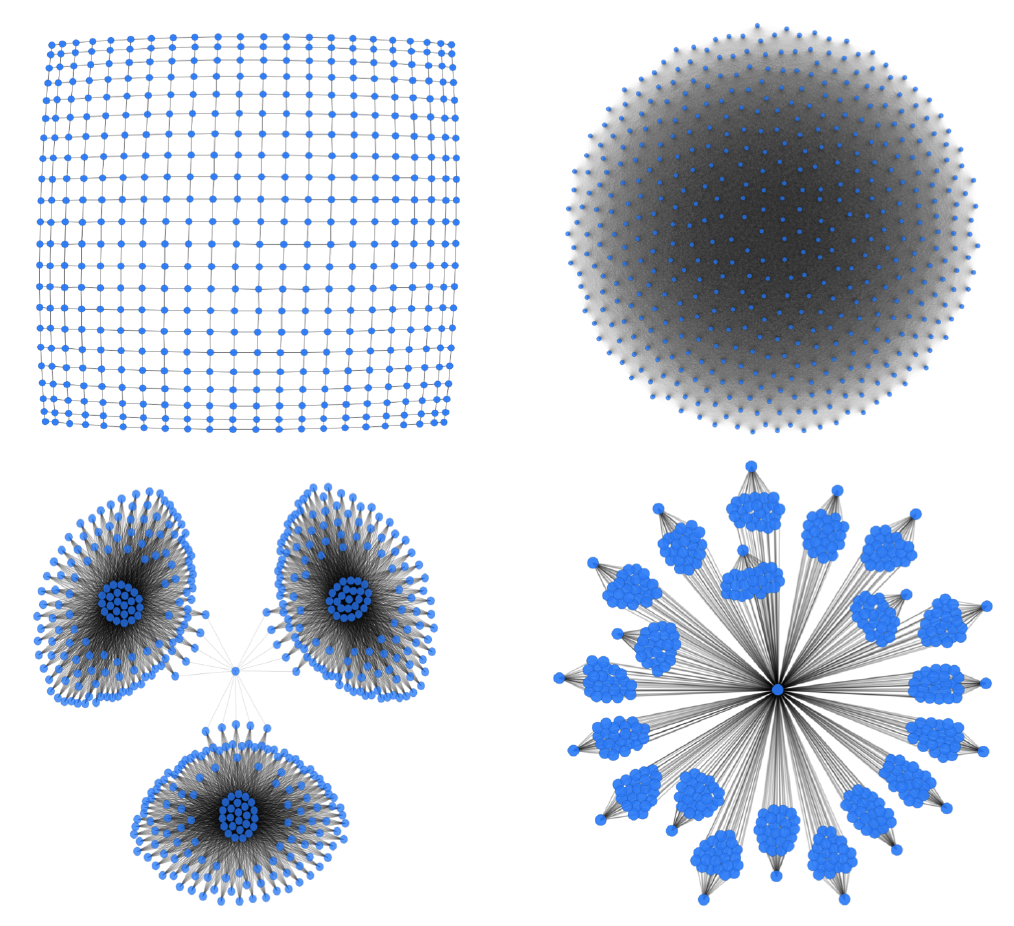}
\caption{The idealised networks described in the text, including a square lattice (upper left), a fully connected network (upper right), a ``meta-funnel'' network (lower left), and a ``superstar'' network (lower right).}
\label{fig:2}       
\end{center}
\end{figure*}

The four network structures, represented in Fig.~\ref{fig:2}, can intuitively be ranked from least to most hierarchical as follows: (1) the square lattice structure, (2) the fully connected network, (3) the metafunnel, and finally (4) the superstar. They are all undirected, and have approximately 475 nodes each, although the precise number depends on the parameterisation of the network in question, which we discuss in Sect.~\ref{sec:2.2}.

The square lattice is a network arrangement that is determined by one parameter: n. The nodes are placed on an n-by-n square grid and each node is connected to the nodes directly above and below it, and to the nodes directly to the left and right of it (there is no wrap around the sides of the grid). The number of nodes in such a network is given by $n^2$. We have set $n = 22$, thus resulting in 484 nodes.

The fully connected network is simply a network in which every node is connected with every other node. It is the special case that corresponds to the neutral model of Bentley {\it et al.} (2011a, 2011b) when the memory parameter is set to one.  The number of nodes was set at $n = 475$.

The metafunnel is a network arrangement that is determined by three parameters: $k$, steps, and $g$. There is a ``central'' node that is connected to $g$ groups of nodes, each group consisting of $k$ nodes. Every node in some group is then connected to the same $k^2$ nodes outward from the central node, and then those nodes are themselves connected to the same $k^3$ nodes, and so on. That is, any node that is $n$ steps away from the central node is connected to $k^{n+1}$ nodes.  We chose $k = 5$ over 3 steps with $g = 3$, which results in 466 nodes.

The superstar is a network arrangement that is determined by two parameters: $s$ and $h$. There is a central node that is connected to every node. The nodes are divided into $h$ groups, each of size $s$. In each group there is a dominant node that is connected to every other node in the group, while all the other nodes within the group are not directly connected to each other. The number of nodes in such a network is given by $1 + sh$. The parameters were set to $s = 24$, and $h = 20$, thus resulting in 481 nodes.

\section{Results}
\label{sec:3}

The expected number of choices at the end of a model run is approximated by $n(1+\mu(T-1))$, so for $\mu = 0.0025$, for example, there are almost 3,000 alternatives from which to choose whilst with $\mu = 0.05$, this number rises to some 50,000.   But even with very low values of $\mu$, a large number of choices exist.

\begin{figure*}
 \includegraphics[width=1\textwidth]{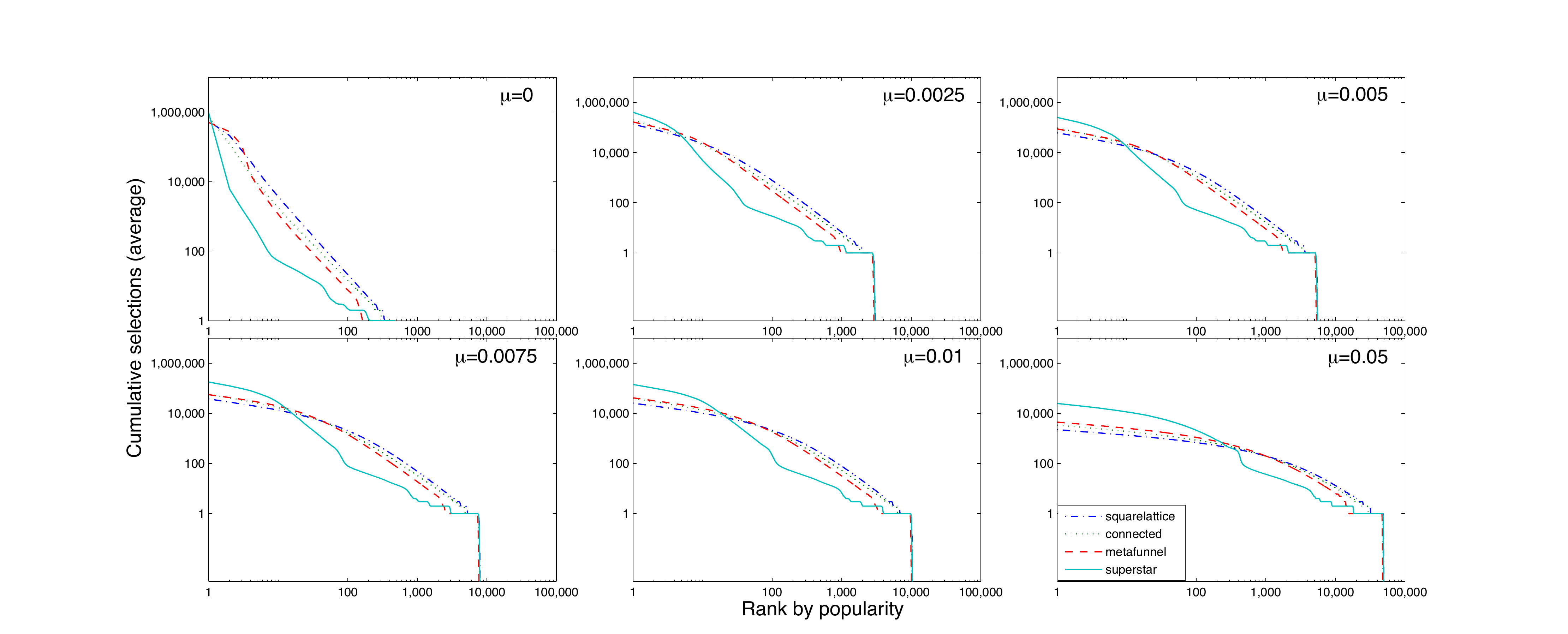}
\caption{The resulting rank-size distribution of choices after 2,000 steps of the model for different value of $\mu$, the innovation parameter.}
\label{fig:3}       
\end{figure*}

Fig.~\ref{fig:3} shows the resulting rank-size distribution of choices after 2,000 steps of the model for different value of $\mu$, the innovation parameter.   We sum the results across all the individual solutions of the model and divide by the total number of solutions.   The most frequently selected alternative has rank 1, the second most frequent rank 2, and so on.

It is known (Neiman, 1995, Bentley {\it et al.} 2011b) that as the value of the innovation parameter, $\mu$, approaches zero, the outcome approaches a winner-take-all distribution.  A small number of alternatives are selected by large numbers of agents, with most choices being selected by few.  The scale on the left hand axes of Fig.~\ref{fig:3} confirms this, with, for example, the top value of the chart when $\mu= 0.0025$ being a whole order of magnitude greater than when $\mu= 0.05$.

All the resulting distributions are highly right-skewed and non-Gaussian.  In general, the more hierarchical is the network structure, the more unequal is the resulting distribution amongst the alternatives which are selected.  The alternatives which are selected most frequently in the case of the superstar network, for example, are chosen considerably more times than the highest ranked with the other distributions. 

The scale of the charts might tend to deceive, but close inspection shows that the number of times the highest ranked alternative is selected with the superstar network is typically an order of magnitude greater than with the other network structures.  Table~\ref{tab:1}illustrates this point.

\begin{table*}
\caption{The number of times the top ranked (by popularity) alternative is selected after 2000 periods (average over 500 runs).}
\label{tab:1}       
\begin{tabular}{lrrrr}
\hline\noalign{\smallskip}
$\mu$&squarelattice&connected&metafunnel&superstar\\
\noalign{\smallskip}\hline\noalign{\smallskip}
0&553422&733172&502469&951377\\
0.0025&134953&197937&165195&401787\\
0.005&61058&94358&86592&250758\\
0.0075&37412&57677&55593&176312\\
0.01&25396&38612&41468&141457\\
0.05&2250&3386&4432&24706\\
\noalign{\smallskip}\hline
\end{tabular}
\end{table*}

\begin{figure*}
 \includegraphics[width=1\textwidth]{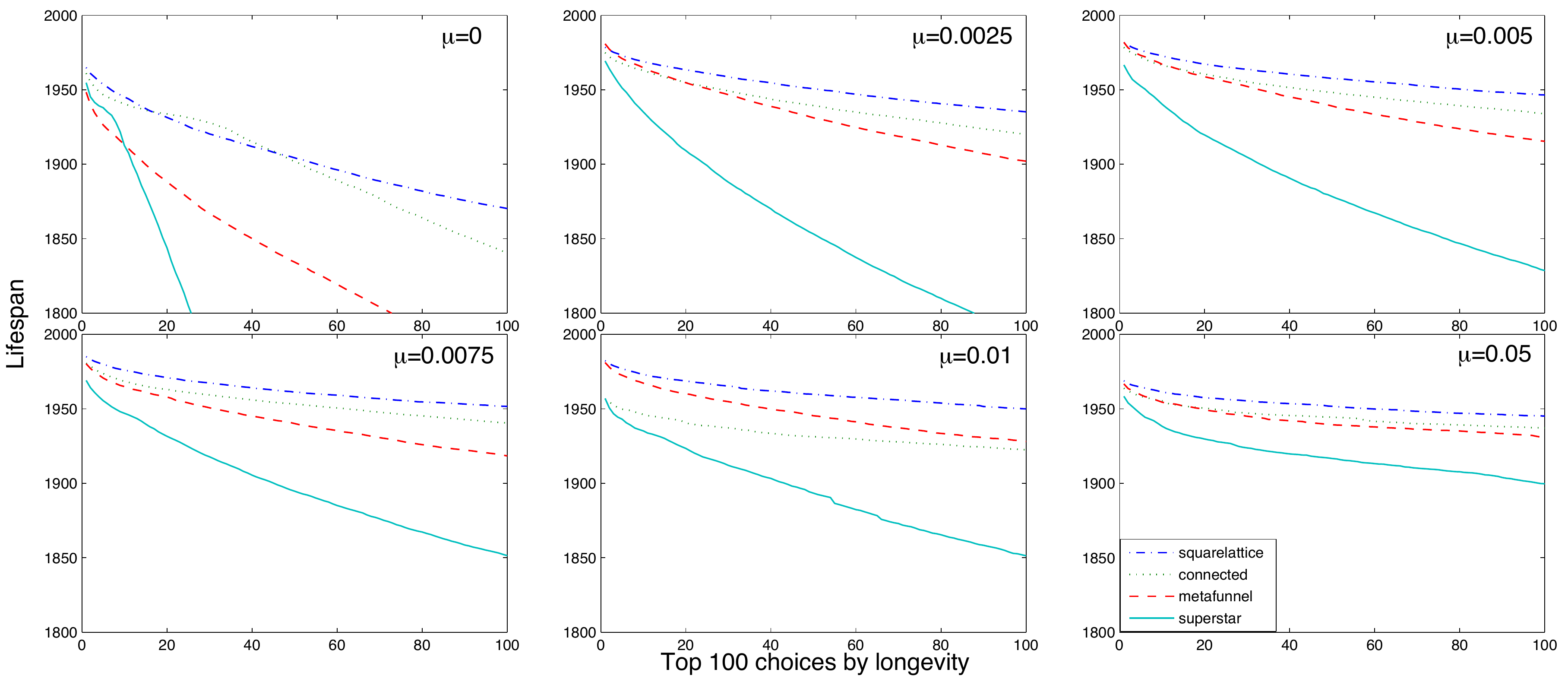}
\caption{Distributions of the lifespans of the top 100 longest-surviving choices, for different networks and different value of $\mu$.}
\label{fig:4}       
\end{figure*}

Fig.~\ref{fig:4} shows the lifespans of the top 100 longest-surviving choices (where the lifespan of a choice is the number of periods for which the choice is selected by some node, starting from the period at which it is first selected to the period after which no agents select it). In general the less hierarchical the network, the more egalitarian the outcome tends to be; that is, the more similar the lifespan of the longest and 100th longest choices tends to be.  

\begin{figure*}
 \includegraphics[width=1\textwidth]{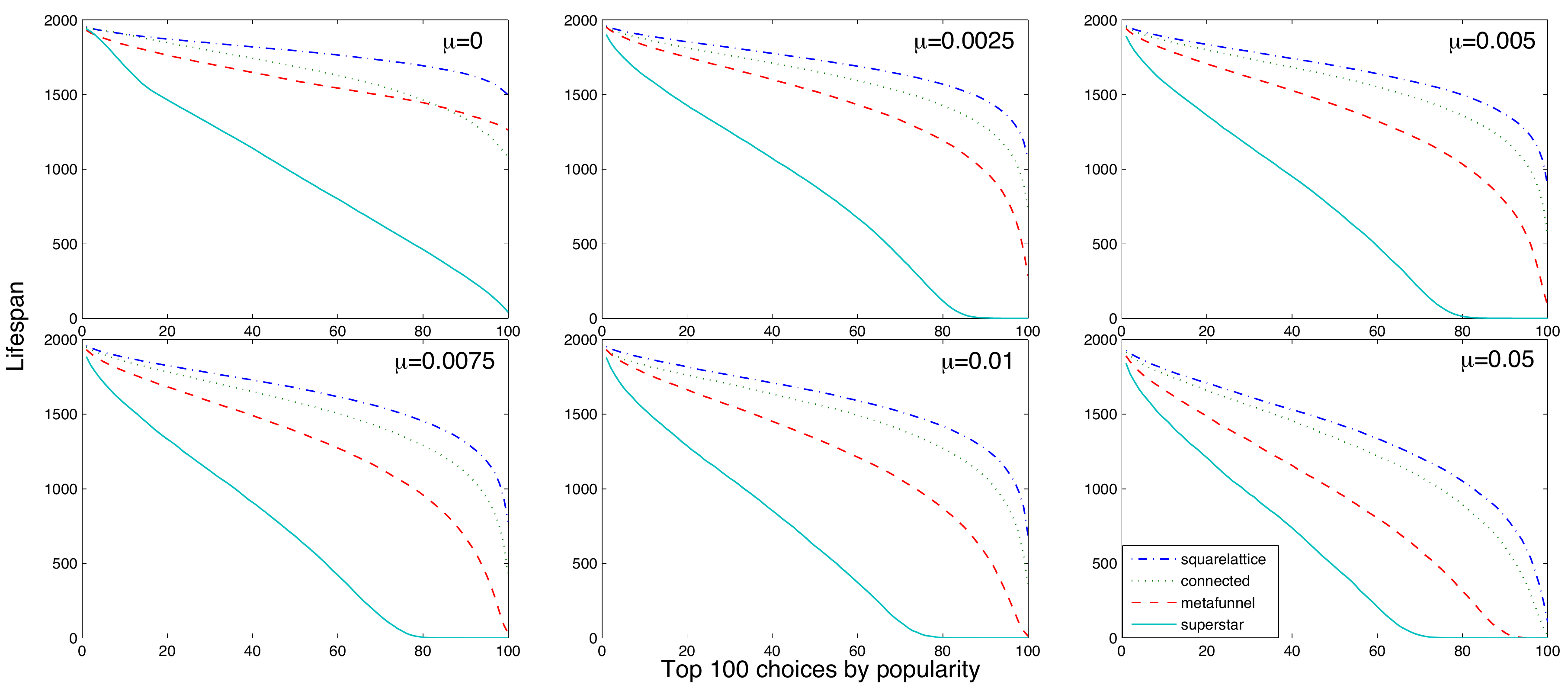}
\caption{Distributions of the lifespans of the top 100 most popular choices, for different networks and different value of $\mu$.}
\label{fig:5}       
\end{figure*}

Fig.~\ref{fig:5} shows the lifespans of the top 100 most popular choices.  In other words, the 100 with the highest number of being selected over all the solutions. It is important to note that in all network structures, the top 100 longest- surviving choices are often distinct from the top 100 most popular choices. In fact, Table~\ref{tab:2} shows that the intersection of these two sets is often quite narrow, where, at best, approximately 20 choices are among both the top 100 most popular and top 100 longest--surviving.

\begin{table*}
\caption{Average number of choices that are among the 100 most popular {\it and} the top 100 most long-lived}
\label{tab:2}       
\begin{tabular}{lllllll}
\hline\noalign{\smallskip}
  & $\mu=0$ & $\mu=0.0025$ & $\mu=0.005$& $\mu=0.0075$& $\mu=0.01$& $\mu=0.05$\\
\noalign{\smallskip}\hline\noalign{\smallskip}
Square lattice &20.51&3.33&1.86&1.4&0.99&0.24 \\
Connected &21.12&3.55&2.00&1.36&1.00&0.22 \\
Metafunnel &21.09&3.47&1.92&1.34&0.99&0.17 \\
Superstar &22.89&3.60&2.00&1.39&1.11&0.24 \\
\noalign{\smallskip}\hline
\end{tabular}
\end{table*}

Finally, Table~\ref{tab:3} shows that average number of ``active'' choices per period in each of the networks - where a choice is said to be ÒactiveÓ in some period if it is being selected by at least 5 nodes in that period. Clearly, the average number of ``active'' choices is increasing in the experimentation parameter $\mu$ for each network, and for a given $\mu$ it appears to decrease the more hierarchical the network is.

\begin{table*}
\caption{Average number of active curves per period}
\label{tab:3}       
\begin{tabular}{lllllll}
\hline\noalign{\smallskip}
  & $\mu=0$ & $\mu=0.0025$ & $\mu=0.005$& $\mu=0.0075$& $\mu=0.01$& $\mu=0.05$\\
\noalign{\smallskip}\hline\noalign{\smallskip}
Square lattice &4.39&13.62&19.17&22.96&25.95&32.90 \\
Connected &2.73&9.00&13.25&16.49&19.10&31.04 \\
Metafunnel &3.01&7.23&10.15&12.45&14.38&28.10 \\
Superstar &1.09&1.65&2.17&2.67&3.13&8.33 \\
\noalign{\smallskip}\hline
\end{tabular}
\end{table*}

\section{Discussion}
\label{sec:4}

It might be imagined that if it is not possible for agents to distinguish between the attributes of alternative choices, so that it is as if no one alternative is superior to any other, that the resulting distribution of the number of times each alternative is selected will be uniformly distributed.  However, this is most decidedly not the case.  We observe marked right--skewed non--Gaussian outcomes, as in the real world of social network markets, in which decisive `winners' emerge.  

Rank-size distributions become steeper the more hierarchical the network structure, approaching a winner--take--all distribution in the case of the superstar network. Lifespans are shorter in more hierarchical networks, but whenever there is a winner, it grows much larger in more hierarchical structures than in less hierarchical ones.  Our results may predict consequences for abrupt, `tipping points' in network structure, such as that discovered by Achlioptas {\it et al.} (2009) for sudden coalescence into sparse hierarchical networks, as has been observed for Wikipedia (Bounova 2009). 

We can think of the relation between these results and the characteristics of the network structures. For example, we are led to the conclusion that choices spread quickly and to a larger fraction of the population not in better connected networks (that is, the fully connected network), but in more hierarchical ones. This is because there is more `noise' in better connected networks -- captured by the number of active choices per period.  So adding connections does not necessarily help in improving the speed or scale at which a choice spreads.  This is essentially what the meso level structure (Foster and Potts 2006) does indeed suggest: namely that there is a lot of choice heuristics and value embedded in institutions

Note that neither the average degree nor the degree variance in the networks explains our results. As argued above, a higher average degree (such as in the fully connected network) does not lead to faster or wider spread. Similarly, the degree variance is higher in the square lattice than in the fully connected network, so this cannot fully capture our intuitive hierarchical ranking.  However, spread is related to degree skewness. That is, spread does seem to be larger and faster in networks where there are relatively few but very well-connected nodes (and the rest -- the majority -- are not very well connected).

Overall, the results are shaped strongly by the features of the networks.  This supports the argument of Foster and Potts (2006) that in social network markets, it is the meso--level structure ({\it i.e.} population and system level) which essentially determines the features which are observed at the macro-level (non--Gaussian rank--size distributions, turnover in rankings through time).

\section{Conclusion}
\label{sec:5}

Social network markets pose challenges to economic theorists, as described by Potts et al. (2008).  We have used an evolutionary perspective to try to develop more insights into the nature of these markets. We find that that a clear winner emerges in social network markets even when, by assumption, there is no inherent superiority of any of the alternative choices available to agents.    The outcome becomes more `winner-take-all' the more hierarchical is the network structure.

The stylised facts of social network markets can be obtained from a model in which, by deliberate assumption, no alternative is superior to any of the others.  This may seem a surprising conclusion.  However, even a brief acquaintance with the most popular picture on Flickr or the most popular videos on YouTube will, on most days, offer powerful casual empirical support for this conclusion.

Lieberman et al. (2005) examined the relationship between network structure and the propensity of an alternative which, by assumption, was superior, to be selected by large numbers of agents.  They showed that choices spread more rapidly and widely in more hierarchical networks.   

Our paper demonstrates that this seems to be a very general principle of social network markets.  It is a feature of such markets regardless of whether agents are able to distinguish between the objective attributes of competing alternatives. These results confirm that choices spread more rapidly and widely in more hierarchical social networks. A clear winner emerges in hierarchical networks even without any inherent superiority.  



\begin{thebibliography}{}

\bibitem[]{Adler_1985}Adler, M. (1985). Stardom and talent. {\it American Economic Review}, 75, 208--212.
\bibitem{Adler_2006}Adler, M. (2006). Stardom and talent. In Ginsburgh, V. A. \& Throsby, D. (Eds.), {\it Handbook of the economics of art and culture}, vol. 1 (pp. 895--906). North Holland: Elsevier
\bibitem{Achlioptas_etal_2009}Achlioptas, D., D'Souza, R. M., \& Spencer, J. (2009). Explosive percolation in random networks. {\it Science}, 323, 1453--1455.
\bibitem{Akerlof_1970}Akerlof, G. A. (1970). The market for ``lemons'': Quality uncertainty and the market mechanism. {\it Quarterly Journal of Economics}, 84, 488--500
\bibitem{Antonelli_2011}Antonelli, C. ed., (2011). {\it Handbook on the economic complexity of technological change}. Cheltenham: Edward Elgar.
\bibitem{Arthur_1989}Arthur, W. B. (1989). Competing technologies increasing returns and lock-in by historical events. {\it Economic Journal}, 99, 116--131.
\bibitem{Asch_1955}Asch, S. (1955). Opinions and social pressure. {\it Scientific American}, 193, 31--35
\bibitem{Baumeister_Tierney_2011}Baumeister, R. F., \& Tierney, J. (2011). {\it Willpower}. New York: Penguin.
\bibitem{Beck_2007}Beck, J. (2007). The sale effect of word of mouth: A model for creative goods and estimation for novels. {\it Journal of Cultural Economics}, 31(1), 5--23.
Becker, G., Murphy, K. (1988) A theory of rational addiction. {\it Journal of Political Economy}, 96, 675--700
\bibitem{Beinhocker_2007}Beinhocker, E. (2007). {\it The origin of wealth}. Cambridge, MA: Harvard Business School Press.
\bibitem{Bentley_2008}Bentley R. A. (2008). Random drift versus selection in academic vocabulary. {\it PLoS ONE}, 3(8), e3057.
\bibitem{Bentley_Ormerod_2008}Bentley, R. A., \& Ormerod, P. (2008). Tradition and fashion in consumer choice. {\it Scottish Journal of Political Economy}, 56, 371--381.
\bibitem{Bentley_etal_2011a}Bentley, R. A., Ormerod, P. \& Shennan, S. J. (2011). Population-level neutral model already explains linguistic patterns. {\it Proceedings of the Royal Society B}, 278, 1770--1772.
\bibitem{Bentley_etal_2011b}Bentley R. A., Ormerod, P., \& Batty, M. (2011a). Evolving social influence in large populations. {\it Behavioral Ecology and Sociobiology}, 65, 537--546.
\bibitem{Bentley_etal_2011c}Bentley, R. A., OÕBrien, M. J., \& Ormerod, P. (2011b). Quality versus mere popularity: A conceptual map for understanding human behavior. {\it Mind \& Society}, 10(2), 181--191.
\bibitem{Bounova_2009}Bounova, G. (2009). {\it Topological evolution of networks}. PhD Thesis, M.I.T. 
\bibitem{Brock_Durlauf_1999}Brock, W. A. \&  Durlauf, S. N. (1999) A formal model of theory choice in science. {\it Economic Theory}, 14, 113--130.
\bibitem{Brock_Durlauf_2001}Brock W. A.,  Durlauf, S. N. (2001) Interactions-based models. In Heckman J. J. \&  Leamer E. (Eds.),  {\it Handbook of econometrics} (pp. 3297--3380). Amsterdam: Elsevier Science. 
\bibitem{Brock_Durlauf_2010}Brock W. A.,  \& Durlauf, S. N. (2010) Adoption curves and social interactions. {\it Journal of the European Economic Association} 8: 232--251.
\bibitem{Christakis_Fowler_2007}Christakis, N. A. \& Fowler, J. H. (2007). The spread of obesity in a large social network over 32 years. {\it New England Journal of Medicine}, 357, 370--379.
\bibitem{DeVany_Walls_1996}De Vany, A., \& Walls, W. D. (1996). Bose-Einstein dynamics and adaptive contracting in the motion picture industry. {\it Economic Journal, Royal Economic Society}, 106(439), 1493--1514.
\bibitem{DeVany_2004}De Vany, A. (2004). {\it Hollywood economics}. London: Routledge.
\bibitem{Diederen_etal_2003}Diederen, P., van Meijl, H., \& Wolters, A. (2003). Modernisation in agriculture: What makes a farmer adopt an innovation? {\it International Journal of Agricultural Resources, Governance and Ecology}, 2, 328--342.
\bibitem{Earls_Potts_2004}Earls, P., \& Potts, J. (2004). The market for preferences. {\it Cambridge Journal of Economics}, 28, 619--633.
\bibitem{Eerkens_2000}Eerkens, J. W. (2000). Practice makes within 5\% of perfect. {\it Current Anthropology}, 41, 663--668.
\bibitem{Ehrenberg_1959}Ehrenberg A. S. C. (1959). The pattern of consumer purchases. {\it Journal of the Royal Statistical Society C}, 8, 26--41.
\bibitem{Foster_Potts_2006}Foster, J., \& Potts, J. (2006). Complexity, networks and the importance of demand and consumption in economic evolution. In M. McKelvey \& M. Holman (Eds.), {\it Flexibility and stability in economic transformation}. Oxford: Oxford University Press.
\bibitem{Godart_Mears_2009}Godart, F., \& Mears, A. (2009). How do cultural producers make creative decisions? {\it Social Forces}, 88, 671--692.
\bibitem{Goodhardt_etal_1984}Goodhardt, G. J., Ehrenberg, A. S. C., \& Chatfield, C. (1984). The Dirichlet: a comprehensive model of buying behavior. {\it Journal of the Royal Statistical Society A}, 147, 621--655.
\bibitem{Hahn_2008}Hahn M. W. (2008). Toward a selection theory of molecular evolution. {\it Evolution}, 62, 255--265. 
\bibitem{Hahn_Bentley_2003}Hahn M. W., \& Bentley, R. A. (2003). Drift as a mechanism for cultural change: an example from baby names. {\it Proceedings of the Royal Society B}, 270, S1--S4. 
\bibitem{Kahneman_2011}Kahneman, D. (2011). {\it Thinking, fast and slow}. New York: Farrar, Strauss and Giroux.
\bibitem{Kretschmer_etal_1999}Kretschmer, M., Klimis, G., \& Choi, C. (1999). Increasing returns and social contagion in cultural industries. {\it British Journal of Management}, 10(1), 61--72.
\bibitem{Ijiri_Simon_1964}Ijiri Y., \& Simon, H. A. (1964). Business firm growth and size. {\it American Economic Review}, 54, 77--89.
\bibitem{Larsen_1961}Larsen, O. N. (1961). Innovators and early adopters of television. {\it Sociological Inquiry}, 32, 16--33.
\bibitem{Lieberman_etal_2005}Lieberman, E., Hauert, C., \& Nowak, M. A. (2005). {\it Evolutionary dynamics on graphs. Nature}, 433, 312--316.
\bibitem{Moscovici_etal_1969}Moscovici, S., Lage, E. \& Naffrechoux, M. (1969). Influences of a consistent minority on the responses of a majority in a colour perception task. {\it Sociometry}, 32,Ê365--368.
\bibitem{Neiman_1995}Neiman, F. D. (1995). Stylistic variation in evolutionary perspective. {\it American Antiquity}, 60, 7--36.
\bibitem{Ormerod_1998}Ormerod, P., (1998). {\it Butterfly economics}. London: Faber and Faber.
\bibitem{Ormerod_2005}Ormerod, P., (2005). {\it Why most things fail: evolution, extinction and economics}. London: Faber and Faber. 
\bibitem{Ormerod_2007}Ormerod, P. (2007). Extracting deep knowledge from limited information on evolved social networks. {\it Physica A}, 378, 48--52. 
\bibitem{Ormerod_Wiltshire_2009}Ormerod, P. \& Wiltshire, G. (2009). Binge drinking in the UK: A social network phenomenon. {\it Mind and Society}, 8, 135--152.
\bibitem{Ormerod_2012}Ormerod, P. (2012). {\it Positive linking: How networks can revolutionise the world} Faber and Faber, London.
\bibitem{Potts_2006}Potts, J. (2006). How creative are the super-rich? {\it Agenda}, 13(4), 139--150.
\bibitem{Potts_etal_2008}Potts, J., Cunningham, S., Hartley, J., \& Ormerod, P. (2008). Social network markets: a new definition of the creative industries. {\it Journal of Cultural Economics} 32(3), 167--185.
\bibitem{Reali_Griffiths_2009}Reali, F., \& Griffiths, T. L. (2009). Words as alleles: connecting language evolution with Bayesian learners to models of genetic drift. {\it Proceedings of the Royal Society B}, 277, 429--436.
\bibitem{Rivera_etal_2010}Rivera, M. T., Soderstrom, S. B., \& Uzzi, B. (2010) Dynamics of dyads in social networks: Assortative, Relational, and Proximity Mechanisms. {\it Annual Review of Sociology}, 36, 91--115.
\bibitem{Rendell_etal_2010}Rendell, L., Boyd, R., Cownden, D., Enquist, M., Eriksson, K., Feldman, M. W., Fogarty, L., Ghirlanda, S., Lillicrap, T.  \& Laland, K. N. (2010) Why copy others?  Insights from the social learning tournament. {\it Science}, 328, 208--213.
\bibitem{Rogers_1962}Rogers, E. M. (1962). {\it Diffusion of innovations}. New York: Free Press.
\bibitem{Rosen_1981}Rosen, S. (1981). The economics of superstars. {\it American Economic Review}, 71, 845--858.
\bibitem{Salganik_etal_2006}Salganik, M. J., Dodds, P. S., \& Watts, D. J. (2006). Experimental study of inequality and unpredictability in an artificial cultural market. {\it Science}, 311, 854Ð856.
\bibitem{Sela_Berger_2012}Sela, A., \& Berger, J. (2012). Decision quicksand: how trivial choices suck us in. {\it Journal of Consumer Research}, 39, in press.
\bibitem{Schelling_1973}Schelling, T. C. (1973). Hockey helmets, concealed weapons, and daylight saving. {\it Journal of Conflict Resolution}, 17(3), 381Ð428. 
\bibitem{Simon_1955}Simon, H. A. (1955). A behavioral model of rational choice. {\it Quarterly Journal of Economics}, 69, 99Ð118.
\bibitem{Srinivasan_Mason_1986}Srinivasan, V. \& Mason, C. H. (1986). Nonlinear least squares estimation of new product diffusion models. {\it Marketing Science}, 5, 169--178.
\bibitem{Winterhalder_Smith_2000}Winterhalder, B., \& Smith, E. A. (2000). Analyzing adaptive strategies: human behavioral ecology at twenty-five. {\it Evolutionary Anthropology}, 9, 51--72.
\bibitem{Young_2009}Young, H. P. (2009). Innovation Diffusion in Heterogeneous Populations:Contagion, Social Influence, and Social Learning. {\it American Economic Review}, 99, 1899--1924.
\bibitem{Young_2011}Young, H. P. (2011). The dynamics of social innovation. {\it Proceedings of the National Academy of Sciences USA},108, 21285-21291.

\end{thebibliography}
\end{document}